\documentclass[aps,showpacs,twocolumn,preprintnumbers]{revtex4}

\usepackage{amssymb,amsbsy,bm,amsmath}
\usepackage{nohyperref}

\newcommand{\intp}{\ensuremath{\int\frac{d^3 p}{(2\pi)^3}}}

\newcommand{\intko}{\ensuremath{\int^{\infty}_{-\infty} dk_0}}
\newcommand{\intpo}{\ensuremath{\int^{\infty}_{-\infty} dp_0}}
\newcommand{\intqo}{\ensuremath{\int^{\infty}_{-\infty} dq_0}}
\newcommand{\bk}{\ensuremath{\mathbf{k}}}
\newcommand{\bp}{\ensuremath{\mathbf{p}}}
\newcommand{\bq}{\ensuremath{\mathbf{q}}}
\newcommand{\bhk}{\ensuremath{\widehat{\mathbf{k}}}}
\newcommand{\bhp}{\ensuremath{\widehat{\mathbf{p}}}}
\newcommand{\bhq}{\ensuremath{\widehat{\mathbf{q}}}}
\newcommand{\nn}{\nonumber}

\begin{document}

\title{Gauge dependence of the fermion quasiparticle poles in hot gauge
theories}

\author{Shang-Yung Wang}

\altaffiliation[Current Address: ]{Department of Physics and
Astronomy, University of Delaware, Newark, Delaware 19716, USA}

\email{sywang@physics.udel.edu}

\affiliation{Theoretical Division, MS B285, Los Alamos National
Laboratory, Los Alamos, New Mexico 87545, USA}

\date{\today}

\begin{abstract}
The gauge dependence of the complex fermion quasiparticle poles
corresponding to soft collective excitations is studied in hot
gauge theories at one-loop order and next-to-leading order in the
high-temperature expansion, with a view towards going beyond the
leading order hard thermal loops and resummations thereof. We find
that for collective excitations of momenta $k\sim eT$ the
dispersion relations are gauge independent, but the corresponding
damping rates are gauge dependent. For $k\ll eT$ and in $k\to 0$
limit, both the dispersion relations and the damping rates are
found to be gauge dependent. The gauge dependence of the position
of the complex quasiparticle poles signals the need for
resummation. Possible cancellation of the leading gauge dependence
at two-loop order in the case of QED is briefly discussed.
\end{abstract}

\pacs{
11.10.Wx, 
11.15.-q, 
12.38.Cy  
}

\keywords{Gauge dependence, Finite-temperature field theory, Gauge
field theories, Fermion collective excitations}

\preprint{LA-UR-04-3675}

\maketitle

\section{Introduction}

The demonstration of gauge independence of physical quantities is
of fundamental importance in gauge theories at finite temperature.
Interestingly, the main breakthrough in perturbative gauge
theories at high temperature was in fact motivated by the quest
for gauge-independent damping rates of plasma excitations. It was
realized~\cite{htl} that in hot gauge theories the usual
connection between the order of the loop expansion and powers of
the gauge coupling constant is lost, hence contributions of
leading order in the gauge coupling constant arise from every
order in the loop expansion. Resummation to all orders of these
leading, gauge-independent hard thermal loops
(HTLs)~\cite{htl,Frenkel:br,Klimov,Weldon:aq,Weldon:bn,Weldon:1989ys}
is necessary for a consistent calculation. The HTL resummation
program~\cite{htl,Frenkel:br} leads to an effective theory that
systematically includes contributions of different momentum
scales~\cite{Braaten:1995jr}.

It was proved in Ref.~\cite{Kobes:1990xf} that the singularity
structure (i.e., the position of poles and branch singularities)
of gauge boson and fermion propagators are gauge independent when
all contributions of a given order of a systematic expansion
scheme are accounted for. While the position of quasiparticle
poles in the leading order HTL approximation are completely gauge
independent~\cite{lebellac,Kraemmer:2003gd}, nevertheless gauge
dependence will enter at subleading order and gauge-independent
extensions beyond the leading order HTL results are not yet
available. More recently, several
authors~\cite{Arrizabalaga:2002hn,Mottola:2003vx,Carrington:2003ut}
have shown that the truncated on-shell two-particle-irreducible
(2PI) effective action has a controlled gauge dependence, with the
explicit gauge-dependent terms always appearing at higher order.
It would be interesting to study possible cancellation of the
leading gauge dependence in the singularity structure of gauge
boson and fermion propagators beyond the leading order HTL
approximation.

The propagation of fermions in hot and dense matter is of interest
in a wide variety of physically relevant situations. In
ultrarelativistic heavy ion collisions and the formation and
evolution of the quark-gluon plasma, lepton pairs play a very
important role as clean probes of the early, hot stage of the new
state of matter~\cite{Harris:1996zx}. The propagation of quarks
during the nonequilibrium stages of the electroweak phase
transition is conjectured to be an essential ingredient for
baryogenesis at the electroweak scale in (non)supersymmetric
extensions of the standard model~\cite{Riotto:1999yt}. In stellar
astrophysics, electrons and neutrinos play a major role in the
evolution of dense stars such as white dwarfs, neutron stars and
supernovae~\cite{raffelt}.

The goal of this article is to investigate the gauge dependence of
the \emph{complex} fermion quasiparticle poles corresponding to
\emph{soft} collective excitations at one-loop order and
next-to-leading order in the hight-temperature expansion. There
have been investigations of how the leading HTL dispersion
relations of fermions (and the gauge independence thereof) are
affected by retaining nonleading powers of
temperature~\cite{Peshier:dy,Mitra:1999wz}. While it is known that
quasiparticles at finite temperature will in general acquire
thermal widths due to collisional broadening rendering the
quasiparticle poles complex, the previous authors considered
mainly the real parts of the complex quasiparticle poles
(dispersion relations) without discussing the corresponding
imaginary parts (damping rates). In this article we study the
gauge dependence of both the real and imaginary parts of the
complex quasiparticle poles on equal footing. Furthermore, we
focus on the soft fermion collective excitations of momenta
$k\lesssim eT$, where $e\ll 1$ is the gauge coupling constant and
$T$ the temperature. On the one hand, it allows us to fill gap in
the literature where either a numerical analysis with $k\sim T$
was carried out~\cite{Peshier:dy} or the limiting cases of $eT\ll
k\ll T$ and $k=0$ were considered~\cite{Mitra:1999wz}. On the
other hand, $k\lesssim eT$ is the relevant momentum region at the
heart of the HTL resummation program~\cite{htl,Frenkel:br}.
Progress made in the soft momentum region will certainly shed
light on the issue of going beyond the HTL resummation.

The rest of this article is organized as follows. In
Sec.~\ref{SecII} we calculate the real and imaginary parts of the
one-loop fermion self-energy in general covariant gauges and up to
next-to-leading order in the high-temperature expansion. In
Sec.~\ref{SecIII} we study the gauge dependence of the complex
fermion quasiparticle poles corresponding to soft collective
excitations of momenta $k\lesssim eT$. Possible cancellation of
the leading gauge dependence at two-loop order in the case of QED
is briefly discussed. Sec.~\ref{SecIV} presents our conclusions.
In the Appendix we calculate the vacuum contribution to the
fermion self-energy that is neglected in the main text.

\section{One-loop fermion self-energy in covariant gauges}\label{SecII}

We will carry out our perturbative calculations in the
imaginary-time (Matsubara) formalism (ITF) of finite temperature
field theory~\cite{Kraemmer:2003gd,lebellac}. The continuation to
imaginary time that describes the theory at finite temperature $T$
is obtained by replacing $it\to\tau$ with $0\leq\tau\leq\beta
\equiv 1/T$. Contrary to the usual ITF, however, we will
\emph{not} work in Euclidean spacetime but keep the metric tensor
and the Dirac gamma matrices the same as in Minkowski spacetime.

In ITF it proves convenient to work in the spectral representation
of the propagators. The fermion propagator is given by
\begin{equation}
s(i\nu_n,\bk)=\intko\frac{\rho_f(k_0,\bk)}{k_0-i\nu_n},\quad
\nu_n=(2n+1)\pi T,
\end{equation}
where $\rho_f$ is the free Dirac fermion spectral function (zero
chemical potential)~\cite{lebellac}
\begin{equation}
\rho_f(k_0,\bk)=\not\!\!K\,\mathrm{sgn}(k_0)\delta(K^2),\quad
K=(k_0,\bk),
\end{equation}
with sgn($x$) being the sign function. The gauge boson propagator
is given by
\begin{equation}
d^{\mu\nu}(i\omega_n,\bp)=\intpo\frac{\rho^{\mu\nu}(p_0,\bp)}{p_0-i\omega_n},
\quad \omega_n=2n\pi T,
\end{equation}
where $\rho^{\mu\nu}$ is the free gauge boson spectral function in
general covariant gauges~\cite{Weldon:2000br}
\begin{equation}
\rho^{\mu\nu}(p_0,\bp)=\mathrm{sgn}(p_0)\left[-g^{\mu\nu}+(\xi-1)P^\mu
P^\nu \frac{\partial}{\partial p_0^2} \right]\delta(P^2),
\end{equation}
with $P=(p_0,\bp)$. The covariant gauge parameter $\xi$ is defined
in such a way that $\xi=1$ is the Feynman gauge.

Since at one-loop order the fermion self-energy has the same
structure (up to gauge group factors) in both Abelian and
non-Abelian gauge theories, for notational simplicity we will
consider the Abelian case in what follows. The non-Abelian case
can be obtained through the replacement $e^2\to g^2  C_F$, where
$C_F=(N^2-1)/2N$ is the Casimir invariant of the fundamental
representation in SU($N$) gauge theories.

In QED the one-loop fermion self-energy in ITF is given by
\begin{eqnarray}
\Sigma(i\nu_n,\bk)&=&e^2 T\sum_{i\omega_m}\intp \gamma^\mu
s(i\nu_n+i\omega_m,\bq)\gamma^\nu\nn\\
&&\times d_{\mu\nu}(i\omega_m,\bp), \label{Sigma1}
\end{eqnarray}
where $\bq=\bk+\bp$. Upon substituting the fermion and gauge boson
propagators into \eqref{Sigma1}, the sum over the bosonic
Matsubara frequency can be done easily~\cite{lebellac} leading to
\begin{eqnarray}
\Sigma(i\nu_n,\bk)&=&e^2\intp\intpo\intqo\rho_{\mu\nu}(p_0,\bp)\nn\\
&&\times\gamma^\mu\rho_f(q_0,\bq)\gamma^\nu
\frac{n_B(p_0)+n_F(q_0)}{q_0-p_0-i\nu_n},
\end{eqnarray}
where $n_{B,F}(x)=1/(e^{\beta x}\mp 1)$ are the Bose and Fermi
distribution functions, respectively.

After the analytic continuation $i\nu_n\to\omega+0^+$ to arbitrary
frequency $\omega$, the imaginary part of the (retarded)
self-energy can be readily found to be given by
\begin{eqnarray}
\mathrm{Im}\Sigma(\omega,\bk)&=&e^2\pi\intp\intpo
\intqo\rho_{\mu\nu}(p_0,\bp)\nn\\
&&\times\gamma^\mu\rho_f(q_0,\bq)\gamma^\nu[n_B(p_0)+n_F(q_0)]
\nn\\
&&\times\delta(\omega-q_0+p_0).\label{ImSigma1}
\end{eqnarray}
The real part of the self-energy is obtained from the imaginary
one through the dispersive representation
\begin{equation}
\mathrm{Re}\Sigma(\omega,\bk)=\mathrm{PV}\int^{\infty}_{-\infty}
\frac{dk_0}{\pi}\frac{\mathrm{Im}\Sigma(k_0,\bk)}{k_0-\omega},\label{disp}
\end{equation}
where PV denotes the principal value.

For massless fermion, rotational invariance and chiral symmetry
entail that the fermion self-energy in equilibrium can be
parameterized by~\cite{lebellac,Kraemmer:2003gd}
\begin{equation}
\Sigma(\omega,\bk)=\Sigma^{(0)}(\omega,k)\,
\gamma^0+\Sigma^{(1)}(\omega,k)\,\boldsymbol{\gamma}\cdot\bhk,
\end{equation}
where $k\equiv|k|$, $\bhk=\bk/k$, and $\Sigma^{(0)}$ and
$\Sigma^{(1)}$ are scalar functions
\begin{eqnarray}
\Sigma^{(0)}(\omega,k)&=&\frac{1}{4}\mathrm{tr}[\Sigma(\omega,\bk)\,\gamma^0],\nn\\
\Sigma^{(1)}(\omega,k)&=&-\frac{1}{4}\mathrm{tr}[\Sigma(\omega,\bk)\,
\boldsymbol{\gamma}\cdot\bhk].
\end{eqnarray}
Since we are interested in the self-energy calculated in general
covariant gauges, it is convenient to decompose the former into
gauge-independent and -dependent contributions. We write
\begin{equation}
\Sigma(\omega,\bk)=\Sigma_\mathrm{FG}(\omega,\bk)+\Sigma_\xi(\omega,\bk),
\end{equation}
where $\Sigma_\mathrm{FG}$ is the gauge-independent part
calculated in the Feynman gauge (FG) and $\Sigma_\xi$ is the
remaining gauge-dependent part.

\subsection{The imaginary part}

We first study the gauge-independent contribution
$\mathrm{Im}\Sigma_\mathrm{FG}$. Evaluating the Dirac traces and
performing the trivial integrals over $p_0$ and $q_0$ in
\eqref{ImSigma1}, we obtain
\begin{eqnarray}
\mathrm{Im}\Sigma^{(0)}_\mathrm{FG}(\omega,k)&=&e^2\pi\intp
\frac{1}{2p}\{[1+n_B(p)-n_F(q)]\nn\\
&&\times\delta(p+q-\omega)+[n_B(p)+n_F(q)]\nn\\
&&\times\delta(p-q+\omega)\}+(\omega\to-\omega),\nn\\
\mathrm{Im}\Sigma^{(1)}_\mathrm{FG}(\omega,k)&=&-e^2\pi
\intp\frac{\bhk\cdot\bhq}{2p}\{[1+n_B(p)-n_F(q)]\nn\\
&&\times\delta(p+q-\omega)+[n_B(p)+n_F(q)]\nn\\
&&\times\delta(p-q+\omega)\}-(\omega\to-\omega).\label{ImSigmaFG}
\end{eqnarray}
In the above expressions the support of the energy-conserving
delta functions $\delta(p+q-\omega)$ and $\delta(p+q+\omega)$ is
$\omega>k$ and $\omega<k$, respectively, corresponding to the
usual two-particle cuts, while that of $\delta(p-q\mp\omega)$ is
$\omega^2<k^2$ corresponding to the Landau damping cut which is
purely a medium effect~\cite{lebellac,htl}.

The different contributions in \eqref{ImSigmaFG} have a physical
interpretation in terms of (off-shell) scattering processes taking
place in the thermal medium. The terms proportional to
$\delta(p+q-\omega)$ arise from the processes in which a
(timelike) fermion decays into a fermion and a gauge boson
$f^\ast\to f+\gamma$, and those proportional to
$\delta(p-q+\omega)$ originate in the Landau damping process in
which a (spacelike) fermion scatters off a gauge boson in the
medium $f^\ast+\gamma\to f$. Here the off-shell fermion is denoted
by a superscript ``$\ast$''.

In what follows we will neglect the vacuum contribution, i.e., the
terms which do not contain any thermal distribution functions, as
we are interested mainly in the finite-temperature medium effects.
Interested readers can find in the Appendix the calculation for
the vacuum contribution.

The angular integration over $\eta=\bhk\cdot\bhp$ in
\eqref{ImSigmaFG} can be performed analytically by using change of
variables $\eta\to z=q(\eta)$ and the energy-conserving delta
functions. The requirement that the energy-conserving delta
functions must have a nonempty support restricts the range of the
radial integration over $p$ for fixed $k$ and $\omega$. The
remaining integration over $p$ for arbitrary $k$ and $\omega$ is
an involved numerical task~\cite{Peshier:dy} which, however, is
not very useful for the purpose of studying next-to-leading order
corrections. In order to compare with the leading order HTL
results, we here focus on a high-temperature expansion for which
$k,\omega\ll T$ and keep terms in $\mathrm{Im}\Sigma$ to
$\mathcal{O}(T)$ in the Landau damping cut contribution but to
$\mathcal{O}(T^0)$ in the two-particle cut contribution.

A comment here is in order. One would presumably expect that it is
sufficient to keep terms in $\mathrm{Im}\Sigma$ uniformly up to
next-to-leading order in the high-temperature expansion, namely,
terms of order $\mathcal{O}(T)$. Whereas this is correct for
generic $k\sim\omega\ll T$, it is not correct in the limit
$k\ll\omega\ll T$ which is in turn relevant to the effective
fermion mass and the damping rate at rest that is obtained in the
long-wavelength limit ($k\to 0$). As will be seen below, the
$\mathcal{O}(T^0)$ two-particle cut contribution in
$\mathrm{Im}\Sigma$ will give rise to $\mathcal{O}(\ln T)$ term in
$\mathrm{Re}\Sigma$ which will become next-to-leading order in $T$
in the limit $k\ll\omega\ll T$.

Separating the leading HTL contributions that can be calculated
analytically, then in the remaining contributions expanding the
distribution functions in the high-temperature limit and cutting
off the potentially divergent momentum integrations at $T$, we
obtain after some algebra ($k,\omega\ll T$)
\begin{eqnarray}
\mathrm{Im}\Sigma^{(0)}_\mathrm{FG}(\omega,k) &\simeq&\frac{\pi
e^2 T^2}{16k}\left[1 +\frac{2\omega}{\pi^2T}
\ln\left|\frac{\omega+k}{\omega-k}\right|\right]\nn\\
&&\times\theta(k^2-\omega^2)+\frac{e^2 T}{8\pi}
\bigg[\frac{\omega}{k}\ln\left|\frac{\omega+k}{\omega-k}\right|\nn\\
&&-1-\frac{|\omega|}{2T}\bigg]\theta(\omega^2-k^2),\nn\\
\mathrm{Im}\Sigma^{(1)}_\mathrm{FG}(\omega,k)&\simeq& -\frac{\pi
e^2 T^2}{16k}\bigg[\frac{\omega}{k}
+\frac{k}{\pi^2T}\left(1+\frac{\omega^2}{k^2}\right)\nn\\
&&\times\ln\left|\frac{\omega+k}{\omega-k}\right|\bigg]
\theta(k^2-\omega^2)+\frac{e^2 T}{8\pi}\nn\\
&&\times \bigg[\frac{\omega}{k}
-\frac{1}{2}\left(1+\frac{\omega^2}{k^2}\right)
\ln\left|\frac{\omega+k}{\omega-k}\right|\nn\\
&&+\mathrm{sgn}(\omega)\frac{k}{2T}\bigg]
\theta(\omega^2-k^2),\label{ImSigma2}
\end{eqnarray}
where $\theta(x)$ is the Heaviside step function. Three important
features are gleaned from the above expressions: (i) The
gauge-independent part $\mathrm{Im}\Sigma_\mathrm{FG}$ receives
contributions both from above and below the light cone,
corresponding respectively to the two-particle and Landau damping
cuts. (ii) The leading $\mathcal{O}(T^2)$ terms in
$\mathrm{Im}\Sigma_\mathrm{FG}$ are recognized as the HTL results,
which arise solely from the Landau damping
process~\cite{htl,lebellac,Weldon:bn,Weldon:1989ys}. (iii) The
subleading terms are suppressed by inverse powers of $T$ and
originate in the Landau damping as well as in the fermion decay
processes.

Next, we calculate the gauge-dependent contribution
$\mathrm{Im}\Sigma_\xi$. The gauge-dependent part of the free
gauge boson spectral function contains the derivative of the
on-shell delta function $\partial\delta(p_0^2-p^2)/\partial
p_0^2$, thus the integral over $p_0$ can be done using integration
by parts which in turn gives rise to derivatives of the Bose
distribution function $dn_B(p)/dp$ as well as of the
energy-conserving delta functions
$\partial\delta(p-q-\omega)/\partial\omega$, etc. Such structures
are expected to be generic to the gauge-dependent higher loop
contributions linear in $(\xi-1)$.

After some lengthy but straightforward algebra we obtain
\begin{widetext}
\begin{eqnarray}
\mathrm{Im}\Sigma^{(0)}_\xi(\omega,k)&=&(\xi-1)e^2\pi\int\frac{d^3p}{4(2\pi)^3}
\bigg\{(1+\bhp\cdot\bhq)\bigg[[1+n_B(p)-n_F(q)]
\frac{\partial}{\partial\omega}\delta(p+q-\omega)\nn\\
&&-\frac{dn_B(p)}{dp}\delta(p+q-\omega)\bigg]-(1-\bhp\cdot\bhq)
\bigg[[n_B(p)+n_F(q)]\frac{\partial}{\partial\omega}\delta(p-q+\omega)\nn\\
&&+\frac{dn_B(p)}{dp}\delta(p-q+\omega)\bigg]\bigg\}+(\omega\to-\omega),\nn\\
\mathrm{Im}\Sigma^{(1)}_\xi(\omega,k)&=&(\xi-1)e^2\pi\int\frac{d^3p}{4(2\pi)^3}
\bigg\{\frac{(\bhk\cdot\bhp)(\bhp\cdot\bhq)-\bhk\cdot\bhq}{p}
\Big[[1+n_B(p)-n_F(q)]\nn\\
&&\times\delta(p+q-\omega)+[n_B(p)+n_F(q)]
\delta(p-q+\omega)\Big]+\bhk\cdot\bhp\bigg[(1+\bhp\cdot\bhq)\nn\\
&&\times\bigg([1+n_B(p)-n_F(q)]
\frac{\partial}{\partial\omega}\delta(p+q-\omega)
-\frac{dn_B(p)}{dp}\delta(p+q-\omega)\bigg)\nn\\
&&+(1-\bhp\cdot\bhq)\bigg([n_B(p)+n_F(q)]\frac{\partial}{\partial\omega}
\delta(p-q+\omega)
+\frac{dn_B(p)}{dp}\delta(p-q+\omega)\bigg)\bigg]\bigg\}\nn\\
&&-(\omega\to-\omega).\label{ImSigmaXi}
\end{eqnarray}
\end{widetext}
It is noted that the various contributions in \eqref{ImSigmaXi},
being gauge dependent and hence in contrast to those in
\eqref{ImSigmaFG}, will not have a physical interpretation in
terms of the scattering processes taking place in the medium. The
gauge-dependent contributions are expected to cancel in physical
quantities in a consistent calculation that generally requires
resummation of perturbation theory. Clearly such a task is beyond
the scope of this article.

The angular integration in \eqref{ImSigmaXi} can be done
analytically as in the gauge-independent contribution. Again
neglecting the vacuum contribution, we find in the
high-temperature limit the following rather compact expressions
($k,\omega\ll T$)
\begin{eqnarray}
\mathrm{Im}\Sigma^{(0)}_\xi(\omega,k)&\simeq&(\xi-1)\frac{e^2
T}{16\pi}\bigg[1+\frac{\omega}{k}\ln\left|
\frac{\omega+k}{\omega-k}\right|\nn\\
&&-\frac{|\omega|}{T}\theta(\omega^2-k^2)\bigg],\nn\\
\mathrm{Im}\Sigma^{(1)}_\xi(\omega,k)&\simeq&(\xi-1)\frac{e^2
T}{16\pi}\bigg[\frac{\omega}{k}-\frac{1}{2}
\left(1+\frac{\omega^2}{k^2}\right)
\ln\left|\frac{\omega+k}{\omega-k}\right|\nn\\
&&+\mathrm{sgn}(\omega)\frac{k}{T}\theta(\omega^2-k^2)\bigg],
\label{ImSigma3}
\end{eqnarray}
where we have combined part of the contributions from above and
below the light cone by using the identity
$\theta(x)+\theta(-x)=1$. The leading gauge-dependent terms in
$\mathrm{Im}\Sigma_\xi$ are of order $\mathcal{O}(T)$ and the
subleading ones are suppressed by inverse powers of $T$ as is the
case for $\mathrm{Im}\Sigma_\mathrm{FG}$.

\subsection{The real part}

From the above results for the imaginary part of the fermion
self-energy the corresponding real part can be obtained using
\eqref{disp}. In the high-temperature limit that we are interested
in, the upper and lower limits of the integral over $k_0$ in
\eqref{disp} can be cut off at $T$ and $-T$, respectively. The
contributions to $\mathrm{Re}\Sigma$ from the ignored regions of
integration are at most of order $\mathcal{O}(T^0)$, hence are
negligible at the next-to-leading order under consideration. It is
worthy noting that since $\mathrm{Im}\Sigma^{(0)}$
[$\mathrm{Im}\Sigma^{(1)}$] is an even (odd) function of $\omega$
therefore, as a result of \eqref{disp}, $\mathrm{Re}\Sigma^{(0)}$
[$\mathrm{Re}\Sigma^{(1)}$] is an odd (even) function of $\omega$.

In the high-temperature limit we obtain for the gauge-independent
part ($k,\omega\ll T$)
\begin{eqnarray}
\mathrm{Re}\Sigma^{(0)}_\mathrm{FG}(\omega,k)&\simeq&-\frac{e^2
T^2}{16k}\left(1+\frac{2k}{\pi^2T}\right)
\ln\left|\frac{\omega+k}{\omega-k}\right|\nn\\
&&-\frac{e^2\omega}{16\pi^2}\ln\frac{T^2}{|\omega^2-k^2|},\nn\\
\mathrm{Re}\Sigma^{(1)}_\mathrm{FG}(\omega,k)&\simeq& -\frac{e^2
T^2}{8k}\left(1+\frac{2k}{\pi^2T}\right)\left[1
-\frac{\omega}{2k}\ln\left|\frac{\omega+k}{\omega-k}\right|\right]\nn\\
&&+\frac{e^2k}{16\pi^2}\ln\frac{T^2}{|\omega^2-k^2|},\label{ReSigma2}
\end{eqnarray}
where we have kept terms up to $\mathcal{O}(\ln T)$ as remarked
above. This is because the $\mathcal{O}(T)$ term in
$\mathrm{Re}\Sigma^{(0)}$ [$\mathrm{Re}\Sigma^{(1)}$] is
proportional to $k$ ($k^2$) in the limit $k\ll\omega\ll T$,
therefore the putative subleading $\mathcal{O}(\ln T)$ term
becomes next-to-leading order in $T$ in this limit [note that
$\mathrm{Re}\Sigma^{(1)}(\omega,k)$ vanishes identically in the
limit $k\to 0$ due to rotational symmetry].

Again the leading $\mathcal{O}(T^2)$ terms in
$\mathrm{Re}\Sigma_\mathrm{FG}$ are the HTL
results~\cite{Klimov,Weldon:bn,Weldon:1989ys,lebellac,Kraemmer:2003gd}.
The $\mathcal{O}(\ln T)$ terms in \eqref{ReSigma2} arise from the
$\mathcal{O}(T^0)$ terms \emph{above} the light cone in
\eqref{ImSigma2}, thus originating in the region of hard loop
momentum. Such $\ln T$ contributions are not unique in the
one-loop fermion self-energy. Indeed, similar $\ln T$ behavior
that originates also in the hard loop momentum region has been
found in the gauge boson
polarization~\cite{Weldon:aq,Brandt:1991fs} as well as, in
general, the $n$-point vertex function~\cite{Brandt:1997rz} at
one-loop order in high temperature non-Abelian gauge theories.

Similarly we find for the gauge-dependent part ($k,\omega\ll T$)
\begin{eqnarray}
\mathrm{Re}\Sigma^{(0)}_\xi(\omega,k)
&\simeq&-(\xi-1)\frac{e^2\omega}{16\pi^2}
\ln\frac{T^2}{|\omega^2-k^2|},\nn\\
\mathrm{Re}\Sigma^{(1)}_\xi(\omega,k)
&\simeq&(\xi-1)\frac{e^2k}{16\pi^2}
\ln\frac{T^2}{|\omega^2-k^2|}.\label{ReSigma3}
\end{eqnarray}
The $\ln T$ contributions in \eqref{ReSigma3}, like the
gauge-independent ones in \eqref{ReSigma2}, arise from the region
of hard loop momentum as well. Following the argument in
Ref.~\cite{Weldon:bn} based on the ultraviolet divergences in the
absence of distribution functions, one might expect the leading
term in $\mathrm{Re}\Sigma_\xi$ to be linear in $T$. Nevertheless,
explicit calculation shows that the leading term actually goes
like $\ln T$ at high temperature. As we will see momentarily, the
absence of the $\mathcal{O}(T)$ term in $\mathrm{Re}\Sigma_\xi$
have an important consequence in the gauge dependence of fermion
dispersion relations.

Furthermore, upon comparing \eqref{ReSigma2} and \eqref{ReSigma3},
we find that the $\mathcal{O}(\ln T)$ term in
$\mathrm{Re}\Sigma_\xi$ has the same prefactor as that in the
corresponding $\mathrm{Re}\Sigma_\mathrm{FG}$, hence they can be
combined together to yield a single gauge-dependent term
proportional to $\xi$. This, however, is not the case for the
imaginary part of the self-energy, where only some of the terms in
$\mathrm{Im}\Sigma_\xi$ and $\mathrm{Im}\Sigma_\mathrm{FG}$ can be
combined together.

\section{Gauge dependence of the fermion quasiparticle poles}\label{SecIII}

Having calculated the one-loop fermion self-energy at
next-to-leading order in $T$ in general covariant gauges, we now
proceed to study gauge dependence of the complex fermion
quasiparticle poles.

The full (retarded) inverse fermion propagator is given by the
Dyson-Schwinger equation
\begin{equation}
iS^{-1}(\omega,\bk)=\omega\gamma^0-
k\boldsymbol{\gamma}\cdot\bhk+\Sigma(\omega,\bk),\label{DSE}
\end{equation}
where $\Sigma(\omega,\bk)$ is the (retarded) fermion self-energy.
Eq.~\eqref{DSE} can be inverted to
yield~\cite{lebellac,Kraemmer:2003gd}
\begin{equation}
S(\omega,\bk)=\frac{i}{2}\bigg[
\frac{\gamma^0-\boldsymbol{\gamma}\cdot\bhk}{\Delta_+(\omega,k)}+
\frac{\gamma^0+\boldsymbol{\gamma}\cdot\bhk}{\Delta_-(\omega,k)}\bigg],
\label{S}
\end{equation}
where
\begin{equation}
\Delta_\pm(\omega,k)=\omega\mp
k+\Sigma_\pm(\omega,k),\label{Deltapm}
\end{equation}
with
\begin{equation}
\Sigma_\pm(\omega,k)=\Sigma^{(0)}(\omega,k)\pm\Sigma^{(1)}(\omega,k).
\label{SigmaPM}
\end{equation}
The analytic continuation of the fermion propagator to complex
frequency features the following singularities:
\begin{itemize}
\item{\textit{Isolated poles}. Isolated \emph{real} poles of the
fermion propagator correspond to stable quasiparticle excitations,
whereas \emph{complex} poles to unstable excitations (resonances)
with finite widths.} \item{\textit{Branch cuts}. Branch cuts of
the fermion propagator correspond to multiparticle states.}
\end{itemize}
As we are interested in the collective excitation in the medium,
we will consider isolated poles in the rest of this section.

Write $\omega=E-i\gamma$ with $E$ and $\gamma$ real. In the narrow
width approximation for which $\gamma\ll E$, the equations that
determine the position of the complex poles are given
by~\cite{Wang:1999mb}
\begin{gather}
E\mp k +\mathrm{Re}\Sigma_\pm(E,k)=0,\nn\\
\gamma+\mathrm{sgn}(\gamma)Z_\pm(k)\mathrm{Im}\Sigma_\pm(E,k)=0,
\end{gather}
where $Z_\pm(k)$ in the second equation are the residues at the
poles (wave function renormalizations)
\begin{equation}
Z_\pm(k)=\left[1+\frac{\partial\mathrm{Re}\Sigma_\pm(\omega,k)}
{\partial\omega}\right]^{-1}_{\omega=E}.\label{Zpm}
\end{equation}
If the product of the residue at the pole and the imaginary part
of the self-energy on the quasiparticle mass shell
$Z_\pm(k)\mathrm{Im}\Sigma_\pm(E,k)$ is negative, there are two
complex poles conjugate to each other in the physical sheet
corresponding to a growing and a decaying solution, i.e., an
instability. On the other hand, if the product is positive there
is no complex pole in the physical sheet, the pole has moved off
into the unphysical (second or higher) sheet. In this case the
spectral function features a Breit-Wigner shape resonance with a
width given by the damping rates
\begin{equation}
\gamma(k)=Z_\pm(k)\mathrm{Im}\Sigma_\pm(E,k),
\end{equation}
which in general determines an exponential falloff
$e^{-\gamma(k)\,t}$ of the fermion propagator in real time. It is
worth noting that numerical results reveal that at next-to-leading
order the product $Z_\pm(k)\mathrm{Im}\Sigma_\pm(E,k)$ is positive
in the Feynman gauge.

From the one-loop fermion self-energy calculated above, we recover
at leading order in $T$ [i.e., $\mathcal{O}(T^2)$] the celebrated
HTL
results~\cite{Klimov,Weldon:bn,Weldon:1989ys,lebellac,Kraemmer:2003gd}.
There are two branches of \emph{stable} fermion collective
excitations (fermions and the so-called
plasminos~\cite{lebellac,Kraemmer:2003gd}) with positive and
negative helicity to chirality ratios, respectively. Their
dispersion relations are manifestly gauge independent and given by
(for positive energy solutions)
\begin{equation}
E(k)=\left\{
\begin{array}{l}
\omega_{+}(k)\quad\text{fermion},\\
\omega_{-}(k)\quad\text{plasmino}.
\end{array}\right.\label{realpole}
\end{equation}
A plot of the HTL dispersion relations $\omega_\pm(k)$ can be
found in the literature (see, e.g.,
Refs.~\cite{lebellac,Kraemmer:2003gd}). One of the main features
is that the two branches of collective excitations develop a gap
$m=eT/\sqrt{8}$, corresponding to an effective thermal mass that
respects both gauge invariance and chiral
symmetry~\cite{Weldon:1989ys,Klimov,Weldon:bn,lebellac,Kraemmer:2003gd}.

The gauge dependence of the quasiparticle poles at next-to-leading
order must be studied with care because the $\mathcal{O}(T)$ terms
in $\mathrm{Re}\Sigma(\omega,k)$ becomes subleading in the limit
$k\ll\omega\ll T$ as remarked above.

Anticipating that the next-to-leading order correction will not
dramatically change the leading order HTL dispersion relations
$\omega_\pm(k)$ which have a gap of order $eT$, we will first
consider the case for which $k\sim eT$. As is clear from the above
results, for generic $k,\omega\sim eT$ the next-to-leading order
correction to $\mathrm{Re}\Sigma(\omega,k)$ is of order $T$ and
gauge independent. Upon substituting $\mathrm{Re}\Sigma$ at this
order into \eqref{Deltapm}, we find that for the collective
excitations of momenta $k\sim eT$ the one-loop dispersion
relations at next-to-leading order in $T$ [i.e., $\mathcal{O}(T)$]
are manifestly gauge independent. This is one of the novel
contributions of this article. Numerical analysis shows that the
dispersion relations at next-to-leading order have similar
features but move slightly above the leading order HTL results in
the same momentum region. The differences increase with increasing
gauge coupling constant, in agreement with the full numerical
result found in Ref.~\cite{Peshier:dy}.

The next-to-leading order correction to $\mathrm{Im}\Sigma$ in the
momentum region $k\sim eT$ is again of order $T$ but gauge
dependent. Whereas at next-to-leading order $\mathrm{Im}\Sigma$
has support \emph{above} the light cone which presumably suggests
finite damping rates $\gamma\sim e^2 T$ (up to the logarithm of
the gauge coupling constant) for the collective excitations at
one-loop order, nevertheless the gauge-dependent contribution to
$\mathrm{Im}\Sigma$ that first appears at this order makes such
interpretations doubtful. This result is not unexpected, as it is
well-known that in high temperature gauge theories a consistent,
gauge-independent calculation of the quasiparticle damping rates
requires a resummation of perturbation theory such as, e.g., the
Braaten-Pisarski HTL resummation program~\cite{htl}.

We next consider the case for which $k\ll eT$. Expanding the
fermion self-energy obtained above in the high-temperature limit
[see \eqref{ImSigma2}, \eqref{ImSigma3}-\eqref{ReSigma3}] in
powers of $k/\omega$ and keeping terms leading in $k/\omega$ and
up to next-to-leading order in $T$, we obtain ($k\ll\omega\sim
eT$)
\begin{eqnarray} \mathrm{Re}\Sigma^{(0)}(\omega,k)&\simeq&
-\frac{e^2 T^2}{8\omega}-\xi\frac{e^2 \omega}{8\pi^2}
\ln\frac{T}{|\omega|},\nn\\
\mathrm{Re}\Sigma^{(1)}(\omega,k)&\simeq&
\frac{k}{\omega}\left[\frac{e^2 T^2}{24\omega}+\xi\frac{e^2
\omega}{8\pi^2} \ln\frac{T}{|\omega|}\right],\label{ReSigma4}
\end{eqnarray}
and
\begin{eqnarray}
\mathrm{Im}\Sigma^{(0)}(\omega,k)&\simeq&
\frac{e^2 T}{8\pi}+(\xi-1)\frac{3e^2 T}{16\pi},\nn\\
\mathrm{Im}\Sigma^{(1)}(\omega,k)&\simeq&
-\frac{k}{\omega}\left[\frac{e^2 T}{6\pi}+(\xi-1)\frac{e^2
T}{12\pi}\right].\label{ImSigma4}
\end{eqnarray}
In the above expressions we have combined the gauge-independent
and -dependent contributions. The next-to-leading order
corrections to $\mathrm{Re}\Sigma$ and $\mathrm{Im}\Sigma$ are of
order $\ln T$ and $T$, respectively, and manifestly
gauge-dependent. Upon substituting the above results into
\eqref{Deltapm}, we find that for the collective excitations of
momenta $k\ll eT$ the one-loop dispersion relations and damping
rates at next-to-leading order in $T$ are gauge dependent.

The effective fermion mass $m$ as well as the damping rate of
collective excitations at rest $\gamma(0)$ at next-to-leading
order can be extracted by taking the long-wavelength limit $k\to
0$. The effective mass is determined by the following equation
($m>0$)
\begin{equation}
m^2=\frac{e^2 T^2}{8}+\xi\frac{e^2
m^2}{8\pi^2}\ln\frac{T}{m},\label{meff}
\end{equation}
which agrees with the result found in the real-time
formalism~\cite{Mitra:1999wz}. Eq.~\eqref{meff} indicates that the
one-loop effective mass at next-to-leading order in $T$ is gauge
dependent and that the gauge-dependent correction to $m$ is of
order $e^4T^2\ln(1/e)$. The damping rate of collective excitations
at rest is of order $e^2T$ but with a gauge-dependent contribution
of the same order.

Before ending this section we discuss briefly possible
cancellation of the leading gauge dependence of the quasiparticle
poles at two-loop order. For simplicity we will consider the $k=0$
case in QED. At two-loop order there are three
one-particle-irreducible (1PI) diagrams contribute to the fermion
self-energy, corresponding to corrections of the vertex, fermion
self-energy and gauge boson polarization. However, because of the
Ward identity for the gauge boson polarization only the first two
will give rise to gauge-dependent contributions. For the purpose
of our discussion here, we only need the two-loop gauge-dependent
contributions linear in $(\xi-1)$. Presumably, the gauge-dependent
contributions quadratic in $(\xi-1)$ will be subleading in $T$.

To extract the leading gauge-dependent contribution at two-loop
order, we focus on diagrams at the level of the propagator rather
than the 1PI self-energy diagrams. Specifically, we consider the
one-particle-reducible (1PR) two-loop self-energy diagram. Using
$\Sigma^{(0)}$ given in \eqref{ReSigma4} and \eqref{ImSigma4}, we
find for the real part (i) the leading gauge-independent
contribution is of order $e^4T^4/\omega^3$ corresponding to what
will be resummed by the HTLs at two-loop order, and (ii) the
leading gauge-dependent contribution is of order
$(e^4T^2/\omega)\ln(T/\omega)$ arising from taking the
gauge-dependent contribution in one of the two loops. Similarly,
for the imaginary part we find both the leading gauge-independent
and -dependent contributions are of order $e^4T^3/\omega^2$.

\emph{Assuming} that the leading gauge-dependent contributions
from the 1PI diagrams have the same behavior as those from the 1PR
one, we obtain in $k\to 0$ limit
\begin{eqnarray}
m^2&=&\frac{e^2 T^2}{8}+\xi\frac{e^2
m^2}{8\pi^2}\ln\frac{T}{m}+a(\xi)\,e^4 T^2\ln\frac{T}{m},\nn\\
\gamma(0)&=&Z(0)\left[\frac{e^2 T}{8\pi}+(\xi-1)\frac{3e^2
T}{16\pi}+b(\xi)\frac{e^4 T^3}{m^2}\right],\label{2loop}
\end{eqnarray}
where $a(\xi)$ and $b(\xi)$ are $\xi$-dependent coefficients yet
to be determined by explicit calculations. To solve for $m$ and
$\gamma(0)$ at two-loop order, one can substitute into terms on
the right-hand side in \eqref{2loop} the leading order result
$m\sim\mathcal{O}(eT)$. \emph{If} $a(\xi)$ and $b(\xi)$ are such
that the gauge-dependent contributions in \eqref{2loop} cancel,
one finds the following gauge-independent results
\begin{equation}
m^2\simeq\frac{e^2
T^2}{8}\left[1+\mathcal{O}\left(e^2\ln\frac{1}{e}\right)\right],\quad
\gamma(0)\simeq\mathcal{O}(e^2 T).
\end{equation}
The above discussion seems to suggest that cancellation of the
leading gauge dependence in $m$ and $\gamma(0)$ at two-loop order
and next-to-leading odder in $T$ is plausible. This situation is
reminiscent of the recent proof that the truncated on-shell 2PI
effective action has a controlled gauge dependence, with the
explicit gauge-dependent terms always appearing at higher
order~\cite{Arrizabalaga:2002hn,Mottola:2003vx,Carrington:2003ut}.

\section{Conclusions}\label{SecIV}

In this article we have studied the gauge dependence of the
fermion quasiparticle poles in hot gauge theories at one-loop
order and next-to-leading order in $T$. We focus on the
quasiparticle poles corresponding to soft collective excitations
of momenta $k\lesssim eT\ll T$, with a view towards going beyond
the leading order HTL results and resummations thereof.

We have calculated the one-loop fermion self-energy (both the real
and imaginary parts) in general covariant gauges up to
next-to-leading order in the high-temperature expansion for which
$k,\omega\ll T$. We find that whereas the next-to-leading order
contributions to the imaginary part behaves like $T$, the behavior
of those to the real part depends on the range of $\omega$ as well
as on their gauge dependence. The next-to-leading order
gauge-independent contribution behaves like $T$ for $k,\omega\sim
eT$, but becomes $\ln T$ for $k\ll\omega\sim eT$. Nevertheless,
the corresponding gauge-dependent contribution always behaves like
$\ln T$ in the high-temperature limit. This analysis allows us to
study in detail the gauge dependence of the complex quasiparticle
poles corresponding to soft collective excitations.

For collective excitations of momenta $k\sim eT$ we find that the
dispersion relations at next-to-leading order in $T$ are gauge
independent. Numerical results shows that these next-to-leading
order dispersion relations have similar features but move slightly
above the leading order HTL ones in the same momentum region.
However, the corresponding damping rates are gauge dependent, thus
rendering the \emph{complex} quasiparticle poles gauge dependent.
For $k\ll eT$ and in the long-wavelength limit $k\to 0$, both the
dispersion relations and the damping rates are gauge dependent.
The gauge dependence of the position of the quasiparticle poles at
one-loop order and next-to-leading order in $T$ signals the need
for resummations of perturbation theory.

We have discussed possible cancellation of gauge dependence at
two-loop order in the case of the effective fermion mass and the
damping rate for collective excitations at rest in QED. Our
analysis suggests that cancellation of the leading gauge
dependence at next-to-leading odder in $T$ is plausible. A
detailed study at two-loop order that allows us to verify such
cancellation in hot gauge theories is the subject of further
investigations.

\acknowledgements

The author thanks E.\ Mottola and G.\ C.\ Nayak
for collaboration during the early stages of this work,
illuminating discussions, and constant interest. He also thanks
M.\ Carrington for discussions and D.\ Boyanovsky for discussions
and carefully reading the manuscript. This work was supported by
the US Department of Energy under contract W-7405-ENG-36.

\appendix

\section{Vacuum contribution to the fermion self-energy}

In this appendix we calculate the vacuum contribution to the
one-loop fermion self-energy for the sake of completeness. As a
by-product, we show that the position of the singularities of the
fermion propagator in vacuum are gauge independent at one-loop
order.

The vacuum contribution to the imaginary part of the self-energy
$\mathrm{Im}\Sigma$ is extracted from \eqref{ImSigmaFG} and
\eqref{ImSigmaXi} by neglecting the thermal distribution
functions. The angular integration can be done analytically in the
same manner as that in the finite-temperature contribution, and
the remaining radial integration over momentum is elementary. We
find
\begin{gather}
\mathrm{Im}\Sigma^\mathrm{vac}_{\mathrm{FG},\pm}(\omega,k)=
\mathrm{sgn}(\omega)\frac{e^2}{16\pi}(\omega\mp k)
\theta(\omega^2-k^2),\nn\\
\mathrm{Im}\Sigma^\mathrm{vac}_{\xi,\pm}(\omega,k)=(\xi-1)\,
\mathrm{sgn}(\omega)\frac{e^2}{16\pi}(\omega\mp
k)\theta(\omega^2-k^2),\nn\\
\end{gather}
where $\Sigma_\pm$ are defined in \eqref{SigmaPM}. We note that
the gauge-independent and -dependent contribution can be combined
together to yield a single gauge-dependent term proportional to
$\xi$. As expected, $\mathrm{Im}\Sigma$ has support only above the
light cone corresponding to the usual two-particle cuts.

The real part obtained from the imaginary one through the
dispersive representation \eqref{disp}, is ultraviolet divergent
and hence has to be regularized by an ultraviolet frequency
cutoff. Introducing counter terms with the renormalization
condition $\mathrm{Re}\Sigma^\mathrm{vac}(\mu)=0$ at some
arbitrary renormalization point $\mu$, we obtain
\begin{equation}
\mathrm{Re}\Sigma^\mathrm{vac}_\pm(\omega,k)= \frac{\xi
e^2}{16\pi^2}(\omega\mp
k)\ln\frac{\mu^2}{|\omega^2-k^2|},\label{ReSigmaVac}
\end{equation}
where the gauge-independent and -dependent contributions have been
combine together.

Upon substituting the above results into \eqref{Deltapm}, we find
that the putative \emph{real} poles of the fermion propagator in
vacuum is determined by
\begin{equation}
(\omega\mp k)\left[1+\frac{\xi e^2}{16\pi^2}
\ln\frac{\mu^2}{|\omega^2-k^2|}\right]=0.
\end{equation}
Because of the infrared logarithmic divergence at threshold
associated with the emission of soft gauge boson, $\omega\to \pm
k$ are no longer isolated poles but the endpoints of logarithmic
branch cuts. Nevertheless, we clear see that the position of the
singularities (in this case logarithmic branch points and branch
cuts) of the fermion propagator in vacuum are gauge independent at
one-loop order.

If one adds the vacuum and the finite-temperature contributions,
one finds that the real part of the vacuum contribution can be
exactly combined with the $\ln T$ term in the real part of the
finite-temperature contribution to yield a single term
proportional to $\ln(T/\mu)$. This generalizes the $k=0$ case
found in Ref.~\cite{Mitra:1999wz} to the case of $k,\omega\ll T$
and is the fermionic counterpart of the results found in
Refs.~\cite{Weldon:aq,Brandt:1991fs,Brandt:1997rz}. However, no
similarly combination can be found in the corresponding imaginary
part of the fermion self-energy.

\end{document}